# Simultaneous Temperature and Acoustic Sensing with Coherent Correlation OTDR

**André Sandmann, Florian Azendorf, Michael Eiselt**
*ADVA Optical Networking SE, Märzenquelle 1-3, 98617 Meiningen, Germany*
*ASandmann@adva.com*

**Abstract:** Superimposed temperature variations and dynamic strain applied through a 400 Hz acoustic signal on a 195 m single-mode fiber section are successfully measured using a coherent correlation optical time domain reflectometry as an interrogator. © 2023 The Author(s)

## 1. Introduction

Distributed optical fiber sensing is of high interest in industry and research since it is capable to monitor several physical parameters in the vicinity of the fiber, while exploiting the benefits of the fiber medium such as long reach and operability in extreme environmental conditions. A challenge is the simultaneous measurement of multiple parameters, such as temperature and strain, when they are superimposed in the same location and therefore change the fiber properties in different ways and on different time scales. Measurements of temperature and dynamic strain at different locations have been demonstrated utilizing a sensor array consisting of ultra-weak fiber Bragg gratings [1]. In order to inscribe such a sensor array, additional devices, e.g. femtosecond laser, various lenses, and drawing machines or a draw tower are required. As a low-cost alternative, a section of standard single-mode fiber can be used as a sensor. The advantage of such fibers is that different scattering processes like Rayleigh, Brillouin and Raman scattering can be exploited to measure different physical parameters. For instance, the frequency dependency of spontaneous Brillouin scattering on temperature and strain has been utilized to measure these parameters simultaneously in different locations [2]. However, the presented strategy is accompanied by increased costs since a frequency shift needs to be realized in the reference path.

In previous work, we reported a temperature measurement by utilizing a correlation optical time domain reflectometry based on direct detection by measuring group delay variations between two reflection points [3]. In this contribution, we demonstrate a novel strategy for simultaneous measurements of temperature and dynamic strain, superimposed at the same location on a single-mode fiber, utilizing coherent correlation optical time domain reflectometry (CC-OTDR) by analyzing the phase information of reflected signal components. It is worth noting that other studies showed simultaneous temperature and strain measurements, but separately applied at different locations of the fiber or by utilizing different fiber types [1,2,4].

## 2. Testbed Setup

In the CC-OTDR testbed setup, compare Fig. 1, the power of the continuous wave laser signal, exhibiting a Lorentzian line-width of less than 100 Hz, is split by a polarization maintaining coupler (PMC). One part is fed to the local oscillator (LO) input of the coherent receiver for polarization diverse homodyne detection, while the other part is modulated with binary phase shift keying (BPSK) using a Mach-Zehnder modulator (MZM). The modulator is driven by an arbitrary waveform generator (AWG). Herein, the transmit signal consists of a 511-bit pseudo-random binary sequence (PRBS) extended by a '-1' symbol, which is transmitted at a rate of 100 MBaud. The PRBS sequence is followed by zero padding, suppressing the output of the modulator while the PRBS test signal propagates in the fiber. The signal is amplified in an erbium doped fiber amplifier (EDFA), and the amplified spontaneous emission noise is limited by an optical bandpass filter (OBF). The resulting signal is fed to the fiber under test (FuT) via an optical circulator and several launch fibers, and the back-scattered and reflected signals are transferred to the signal input (Sig) of the coherent receiver. With a real-time oscilloscope, the four extracted signal components are sampled at a rate of 500 MS/s, stored and further processed offline, comprising a cross-correlation of the received signals with the transmitted sequence.

Different physical contact (PC) connectors are visible as peaks in the return loss trace. The fiber under test is terminated by an open angled physical contact (APC) connector with a dust cap. Rayleigh back-scattering is recognizable and its return loss varies due to coherent fading. A fiber section of 195 m length is put into a temperature-controlled chamber together with a speaker to simultaneously apply temperature changes and an acoustic signal, as seen in the photo in Fig. 1. The reference temperature is measured by a dedicated sensor that is located at the back wall of the chamber. It is worth noting that the temperature chamber includes a fan, which is active when the temperature control of the chamber is switched on. To evaluate the effect of temperature and vibrations on the fiber under test, the propagation phase difference between the reflections at 39 meters and 234 meters are evaluated. Phase variations of the laser over time cancel out, and only the phase noise during the two-way propagation time in the FuT (approx. 1.95 µs, resulting in a phase noise standard deviation of less than 0.05 rad) show as noise in the phase difference. The main part of the phase difference variations over time stem from the refractive index changes due to temperature and strain variations over time.

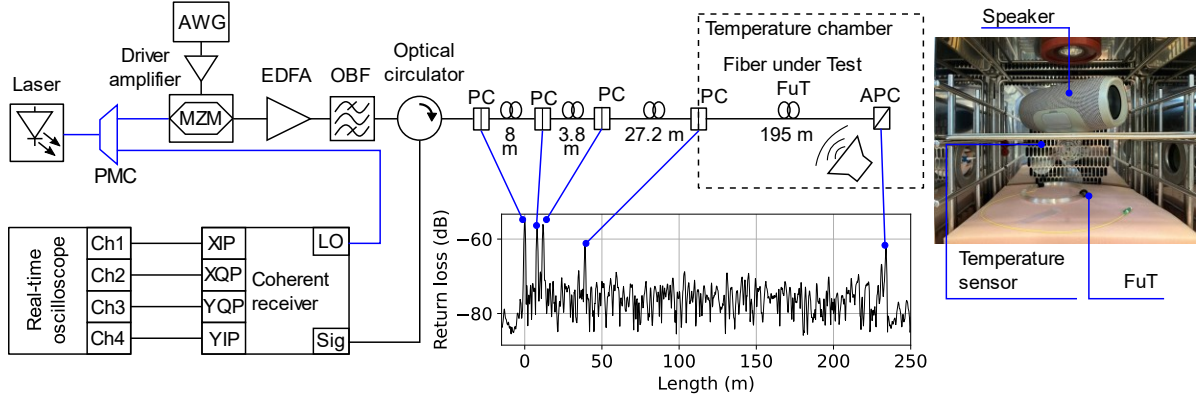

Fig. 1. Coherent correlation OTDR testbed setup with the corresponding fiber fingerprint. The photo shows the placement of speaker and fiber in the temperature chamber.

## 3. Experimental Results

In the temperature chamber, the following profile is applied: In the first 27 seconds, the chamber and hence the fan are switched off at an initial temperature of 30°C. Subsequently, the chamber is switched on with maintaining the initial temperature for the next 27 seconds, which activates the fan. Afterwards, the target temperature is set to 40°C for the next 194 seconds, such that the temperature in the chamber is increased with a maximum change of 0.09 K/s. Finally, the temperature is set to 30°C, resulting in a cooling process with a maximum slope of -0.21 K/s. Simultaneously to the application of temperature change and airflow by the fan, a sinusoidal acoustic signal with 400 Hz is applied.

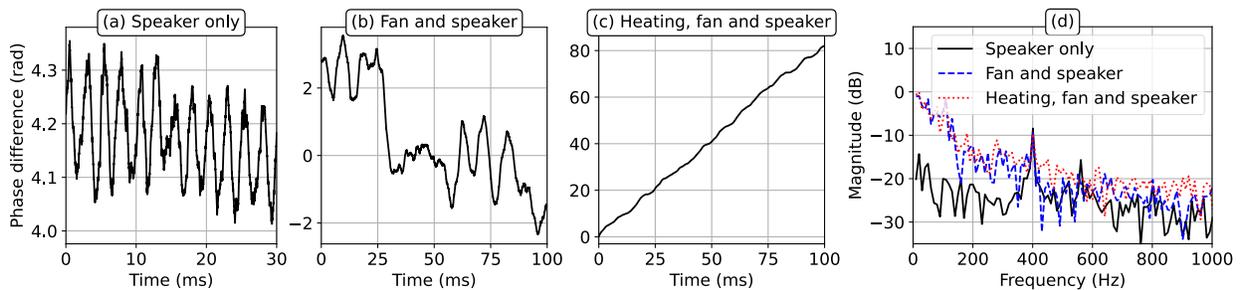

Fig. 2. Phase difference evolutions [(a) to (c)] and slope compensated magnitude spectrum (d) of the 195 m FuT in different test profile stages.

In Fig. 2, the measured evolutions of the phase difference between the reflections at 39 m and 234 m for different time windows and states of the profile are shown. Figure 2(a) depicts the evolution of the phase when the temperature chamber is turned off and just the 400 Hz acoustic wave is applied via the speaker. This sinusoidal signal is clearly visible in the phase difference with a peak-peak amplitude of 0.25 rad as well as in the corresponding spectrum depicted in (d). When the temperature chamber is in the switched on state, the active fan leads to significantly higher changes of the phase due to the airflow-induced strain changes, which is shown in the phase evolution in (b). The increase in spectral magnitude below 400 Hz originates from the activated fan. At the same time, the applied 400 Hz acoustic wave is clearly visible in the spectrum. Considering the heating process, the phase difference increases linearly with 830.5 rad/s in the 100 ms time window as shown in (c). The corresponding slope of the fiber temperature obtained by the phase evaluation is 0.05 K/s and hence the FuT arrangement shows a phase temperature coefficient of 16.6 rad/mK. Consequently, the refractive index variation with temperature results in $10^{-5}$ 1/K, which comes close to values given in [5]. Based on the phase difference slope within each measurement window, the temperature change can be evaluated. By subtracting the constant slope in each time window, fast changes such as the acoustic wave and the variations resulting from the airflow can be separated from the slow temperature change. Alternatively, this can be achieved by considering the spectrum of the phase changes, as indicated in Fig. 2(d). It should be noted that this strategy exploits the property that the applied dynamic strain changes have a higher frequency compared to the temperature changes.

Figure 3 shows the resulting temperature evolution measured with the temperature sensor (black lines) and the obtained temperature from the measured phase evolution of the fiber section (blue lines). For calibrating purposes, a temperature profile was used where only the heating process from 30°C to 40°C was performed, indicated as dotted lines, while the above described heating-and-cooling temperature profile is illustrated with solid lines. The results show that the temperature changes in the fiber core are delayed compared to the temperature sensor reading. Applying a first-order low pass-filter to the temperature sensor data (red lines) shows that a time constant of 73 seconds minimizes the least squares error when comparing to the obtained phase evolution. Furthermore, the temperature in the fiber core reaches an average of 37.5°C in the peak. It needs to be considered that the fiber is air spooled and the parts at the outside of the fiber bundle will have a higher core temperature than the ones in the center of the bundle.

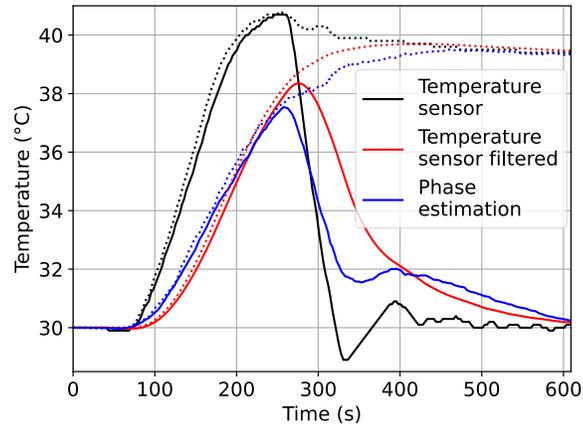

Fig. 3. Temperature measurement results of the 195 m FuT (dotted lines: heating profile 30 to 40°C, solid lines: heating-and-cooling cycle 30 to 40 to 30°C).

## 4. Conclusion

This contribution demonstrates the simultaneous measurement of temperature and dynamic strain on a 195-m fiber section using propagation phase measurements based on a CC-OTDR. By evaluating the slopes of the measured phase differences, the slow temperature process is separated from the fast changes due to an acoustic wave and the airflow from a fan. The analyzed fiber arrangement shows a delayed temperature response that is approximated by a first-order low-pass characteristic with a time constant of 73 seconds. The applied strain from airflow results in an increase of the phase difference spectral magnitudes at frequencies below 400 Hz.

**Acknowledgements**

This work was partially funded by the German Federal Ministry of Education and Research in the framework of the RUBIN project Quantifisens (Project ID 03RU1U071D).